# Current induced spin injection and surface torque
# in ferromagnetic metallic junctions

R.J. Elliott[*], E.M. Epshtein[**], Yu.V. Gulyaev[**], P.E. Zilberman[**][1]

[*] University of Oxford, Department of Physics, Theoretical Physics, Oxford OX1, UK
[**] Institute of Radio Engineering and Electronics, Fryazino, Moscow District, Russia

**Abstract.** Joint influence of two effects, namely, nonequilibrium spin injection by current, and current induced surface torque, on spin-valve type ferromagnetic metallic junctions is considered theoretically. The CPP configuration is assumed. The consideration is based on solving a coupled set of equations of motion for the mobile electron and lattice magnetizations. Boundary conditions are derived from the total magnetization flux continuity condition. A dispersion relation is derived for current dependent spin-wave fluctuations. The fluctuations become unstable under current density exceeding some threshold value (typical value is $1 \times 10^6$ - $3 \times 10^7$ A/cm$^2$). Joint action of the longitudinal spin injection and the torque lowers the instability threshold. The spin injection softens spin wave frequency near the threshold and can pin magnetization at the injecting contact. The pinning rises under the current increasing, so that the appearance of new spin-wave resonance lines can be observed.

## 1. Introduction

Great attention is paid now to investigate features of current flowing through ferromagnetic junctions, i.e., structures with contacting ferromagnetic thin layers. As experiments showed, current can influence substantially the layer magnetic state of such junctions that leads to resistance jumps [1-3], as well as microwave emission [4-6].

The mechanism of the current effect on the junction magnetic state is not clear enough so far. A mechanism was proposed [7] of the current effect on the ferromagnetic layer magnetization $M$ by injection of nonequilibrium longitudinal (i.e., collinear to $M$) spins into the layer. The mechanism theory was formed in [8-10]. The injection creates nonequilibrium carrier spin polarization in the layer. The polarization, in its turn, contributes to the *sd* exchange energy $U_{sd}(\boldsymbol{j})$, as well as to the corresponding *sd* exchange effective field $\boldsymbol{H}_{sd}(\boldsymbol{j})$ dependent on the current density $\boldsymbol{j}$. Above some threshold current density, a reorientation first-order phase transition occurs, as well as an abrupt change in the magnetization vector $\boldsymbol{M}$ direction. Such a current induced magnetization reversal (or current induced switching) leads to resistance jumps that agree well with experimental data [1-3].

On the other hand, another mechanism of the current effect on the ferromagnetic layer state was proposed [11, 12] long before the experimental works [1-6]. According to this mechanism, the transverse (with respect to $\boldsymbol{M}$) electron spin current is to vanish near the interface between two non-collinear ferromagnets. Owing to interaction of the mobile electrons with magnetic lattice (*sd* exchange interaction), a torque appears at the interface that has an effect on the lattice and transfers the lost spin current to it. As a result, the total spin current of the mobile electrons and the lattice remains continuous at the interface.

Besides the current dependent torque, a dissipative torque influences the $\boldsymbol{M}$ vector to restore magnetic equilibrium. With increasing the current density $\boldsymbol{j}$, the dissipation effect is overcome, after all, by the current-induced torque. Then the initial state becomes unstable and the magnetization reversal occurs. As estimates show, such a switching mechanism also agrees with experiments [1-6].

It was assumed in the original works [11, 12] that the transverse electron spin current vanished under ballistic transport conditions. An opposite limiting case, when diffusion transport dominates, is considered in [13, 14]. In the present work, we follow, in this point, the approach of Refs. 11, 12 and assume ballistic transport is valid near the interface. The quantitative validity criteria for this assumption are presented in Section 2.

An important question arises: how are things going in real experiments when both effects coexist, namely, longitudinal spin injection and corresponding current dependent effective field, and current dependent torque at the interface of the magnetic junction. Up to now, these effects were studied separately. Meanwhile, these effects do not only coexist, but influence each other also. Therefore, both effects are to be taken into account simultaneously, in scope of a unified theory, to understand better the experimental situation. Such a theory development is the aim of the present work.

Two comments should be made here. First, the theoretical description of the torque "anatomy" is a rather difficult microscopic quantum problem, as it is seen from the original works [11, 12] and the other works including paper [15] devoted specially to such a description. Undoubtedly, solving that problem is of great interest in itself. However, such an approach would be an excessive complication if we tried to formulate desired unified theory on that base. It seems more reasonable to take advantage of difference between spatial scales of two phenomena indicated, namely, spin injection and torque. The torque is a purely surface effect. It acts near the interface within $d$ distance comparable with the electron Fermi wavelength $\lambda_F \sim 1$ nm. On the other hand, the nonequilibrium longitudinal spins inject inside the layer into much more spin diffusion length, $l \sim 10 - 100$ nm for ferromagnetic metals at room temperature. Consequently, the spin injection is a volume effect, to a considerable extent, in comparison with the torque. Therefore, the spin injection is described below by solving a coupled set of equations for the electron system and the $\textbf{\textit{M}}$ vector in the bulk of the ferromagnetic layer, while the torque is taken into account as a boundary condition for those equations. We show that such a boundary condition can be obtained easy enough from the total spin current conservation requirement.

The second comment concerns a methodical feature of the present work. Unlike from our preceding works [8-10] based on the variation principle of minimal magnetic energy, the problem is solved here by a purely dynamical way based on the linearized Landau – Lifshitz – Gilbert (LLG) equations. In [8-10] we described a reorientation phase transition under spin injection, which corresponded to disappearance of an energy minimum. The surface torque was not taken into account there. As shown below, the torque induced instability has nothing to do with the case of the phase transition and the energy minimum. This instability is due to an energy generation process that overcomes dissipative loss. Therefore, a general approach based on the equations of motion is to be chosen to describe simultaneously both types of instability. As shown earlier [9], both approaches are equivalent with respect to the purely injection mechanism. Such a conclusion is confirmed by calculations in present work, too.

## 2. Model of magnetic junction

Let us consider a spin-valve type magnetic junction with current flowing across the layer interfaces (CPP geometry, see Fig. 1). The junction contains a ferromagnetic layer **1** with fixed orientation of the lattice and mobile electron spins. Such an orientation can be reached if high induced magnetic anisotropy takes place in the layer **1**, and the layer is made of half-metal, in which only one of two spin subbands takes part in conduction [16]. Another ferromagnetic metal layer **2** with magnetization $\textbf{\textit{M}}$ is assumed to contain free spins, so that the magnetization direction can be changed by an external magnetic field $\textbf{\textit{H}}$ or spin-polarized current density $\textbf{\textit{j}}$.

There is a very thin nonmagnetic spacer between the layers **1** and **2** (it is shown as a heavy line in Fig. 1). To close the electric circuit, a nonmagnetic metal layer **3** is necessary.

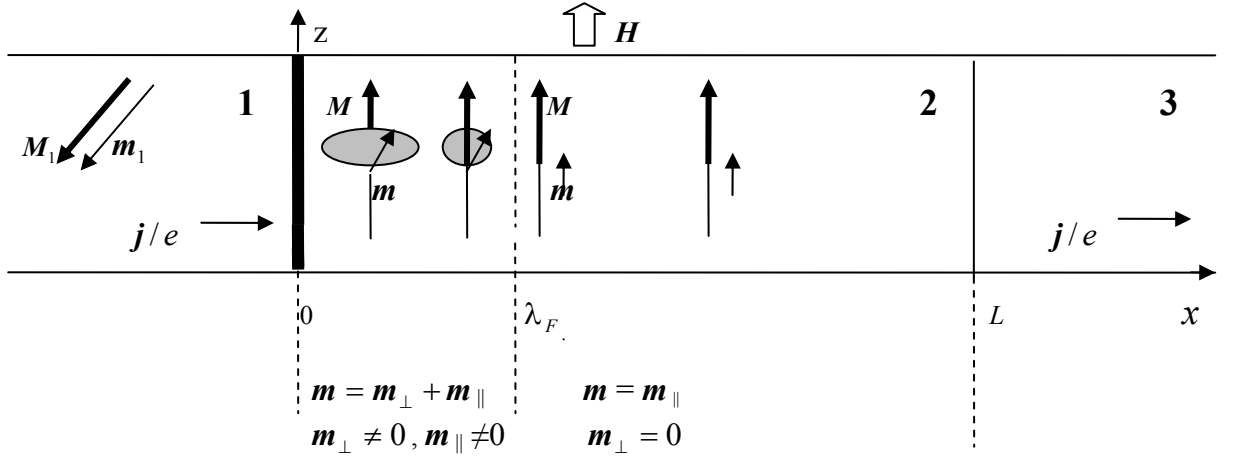

Fig. 1. A scheme of magnetic junction to illustrate processes in the layer **2**. The contacting layers are labeled by **1**, **2**, **3**. The arrows show directions of the following vectors: $\boldsymbol{M}_1$ and $\boldsymbol{m}_1$ are magnetizations in the layer **1**, $\boldsymbol{M}$ and $\boldsymbol{m}$ are magnetizations in the layer **2**, external magnetic field $\boldsymbol{H}$ laying in the junction plane $x = 0$, and electron flux density $\boldsymbol{j}/e$. The vertical dashed lines separate two ranges in the layer **2**. In $0 \le x \le \lambda_F$ range, precession takes place, which is shown with ovals. Vector $\boldsymbol{m}$ has both longitudinal and transversal components, $\boldsymbol{m}_\parallel$ and $\boldsymbol{m}_\perp$, respectively. The precession angle decreases with $x$ increasing. In $x > \lambda_F$ range, the precession vanishes, so that only the longitudinal component remains.

The plane $x = 0$ is the interface between the layers **1** and **2**. Within the layer **1**, the lattice magnetization $\boldsymbol{M}_1$ and the mobile electron magnetization $\boldsymbol{m}_1$ are collinear, while $\boldsymbol{M}_1$ vector is parallel to $x = 0$ plane. $\boldsymbol{M}_1$ and $\boldsymbol{M}$ vectors can make an angle $\chi \ne 0$. Therefore, the electrons transferred to layer **2** by the current appear in a non-stationary quantum state and "walk" between the spin subbands. It corresponds to precession of the mobile electron magnetization $\boldsymbol{m}$ and the lattice magnetization $\boldsymbol{M}$ (see Fig. 1). It was shown in [11, 12] that the spin of each an electron perform a precession with its own initial phase

$$\varphi(d) = \left(k_{x\uparrow} - k_{x\downarrow}\right)d , \qquad (1)$$

which depends on the distance $d$ between the electron and the interface. In (1), $k_{x\uparrow}$ and $k_{x\downarrow}$ are $x$ components of the electron wave vectors in spin-up (along $\boldsymbol{M}$) and spin-down subbands. Because of the electron velocity statistical spread, the expression in the brackets in (1) changes in magnitude from 0 to $2\pi/\lambda_F$ . Therefore, phase (1) is distributed within the interval $0 \le \varphi \le 2\pi$ at $d = \lambda_F$ . Since the transverse (with respect to $\boldsymbol{M}$) component $\boldsymbol{m}_\perp$ of the vector $\boldsymbol{m}$ is a sum of transverse components of individual electrons $\delta \boldsymbol{m}_\perp$, the resultant cancellation takes place with sufficient accuracy, so that

$$\boldsymbol{m}_\perp = \sum \delta \boldsymbol{m}_\perp \to 0 \,. \tag{2}$$

Note, that $l >> \lambda_F$, so that spin relaxation does not affect the validity of condition (2). Besides, this condition was derived under assumption of the ballistic electron motion in $0 \leq x \leq \lambda_F$ range, that is why the $\boldsymbol{k}$ vector was not changed by collisions. This is valid under $l_p > \lambda_F$ condition, where $l_p$ is the momentum relaxation length. Because of $l_p \sim 1 - 10$ nm and $\lambda_F \sim 1$ nm, typically, the ballistic regime condition can fulfil well. The layer **2** is assumed to be extended enough with thickness $L >> \lambda_F, l_p$.

Important conclusions follow from condition (2):

1) The precession in the layer **2** drops at a small distance $d \sim \lambda_F \sim 1$ nm from the layer **1** interface. As a result, the transverse component of the mobile electron magnetization disappears, as well as corresponding spin current. In accordance with [11, 12], it leads to appearance of a torque that acts on the lattice and ensures continuity of the transverse component of the total magnetization ($\boldsymbol{M}_\perp + \boldsymbol{m}_\perp$).

2) At $x > d \sim \lambda_F$, mobile electrons adapt themselves to a new direction of the quantization axis (vector $\boldsymbol{M}$) and occupy new spin subbands. The subband population depends on the current and is not equilibrium one. It is determined by a continuity condition for the longitudinal component of the mobile electron magnetization flux across the interface. The longitudinal component of the flux cannot be changed at small distances $x << l$. This is exactly the spin injection effect that was considered for the first time, apparently, in [17]. Such an effect is similar to the well known charge injection effect in semiconductors [18][2]. In accordance with the typical estimate, $l >> l_p$, the injected electrons motion inside the layer **2** may be considered as a diffusion one.

## 3. Basic equations

Let us consider processes in the layer **2**, which may be described by means of a set of dynamical equations for magnetization vectors $\boldsymbol{M}$ and $\boldsymbol{m}$. The lattice magnetization[3] $\boldsymbol{M}$ obeys LLG equation

---

[2] Spin injection can be realized not only by current, as in [17] or in the present work, but by means of polarized light too (see e.g. [19]). Apparently, the spin injection concept can acquire a great significance in future, comparable with the charge injection concept in semiconductor devices.

[3] We mean a lattice of magnetic ions described within a continuum approximation.

$$\frac{\partial \boldsymbol{M}}{\partial t} = -\gamma\big[\boldsymbol{M}, \boldsymbol{H}_{eff}\big] + \frac{\kappa}{M}\bigg[\boldsymbol{M}, \frac{\partial \boldsymbol{M}}{\partial t}\bigg], \tag{3}$$

where $\gamma$ is the gyromagnetic ratio, $\kappa$ is dimensionless damping constant ($0 < \kappa \ll 1$),

$$\boldsymbol{H}_{eff} = \boldsymbol{H} + \beta(\boldsymbol{Mn})\boldsymbol{n} + A\frac{\partial^2 \boldsymbol{M}}{\partial x^2} + \boldsymbol{H}_d + \boldsymbol{H}_{sd} \tag{4}$$

is effective field, $\beta$ is dimensionless anisotropy constant, $\boldsymbol{n}$ is unit vector along the anisotropy axis, $A$ is the intralattice inhomogeneous exchange constant, $\boldsymbol{H}_d$ is demagnetization field, $\boldsymbol{H}_{sd}$ is $sd$ exchange effective field. The latter takes the form [17]

$$\boldsymbol{H}_{sd}(x,t) = -\frac{\delta U_{sd}}{\delta \boldsymbol{M}(x,t)}, \tag{5}$$

where $\delta/\delta \boldsymbol{M}(x,t)$ is a variation derivative, and $U_{sd}$ is $sd$ exchange energy,

$$U_{sd} = -\alpha \int_0^L \boldsymbol{m}(x',t)\boldsymbol{M}(x',t)dx', \tag{6}$$

parameter $\alpha$ being a dimensionless $sd$ exchange constant (typical estimation is $\alpha \sim 10^4 - 10^6$ [8]). Due to the last term in (4), the motions of $\boldsymbol{M}$ and $\boldsymbol{m}$ vectors appear to be coupled.

The mobile electron magnetization $\boldsymbol{m}$ obeys a general type continuity equation [17, 21]

$$\frac{\partial \boldsymbol{m}}{\partial t} + \frac{\partial \boldsymbol{J}}{\partial x} + \gamma\alpha[\boldsymbol{m}, \boldsymbol{M}] + \frac{\boldsymbol{m} - \overline{\boldsymbol{m}}}{\tau} = 0, \tag{7}$$

where $\tau$ is a time of relaxation to the local equilibrium value $\overline{\boldsymbol{m}} \equiv \overline{m}\hat{\boldsymbol{M}}$, and $\hat{\boldsymbol{M}} \equiv \boldsymbol{M}/M$ is the unit vector, $\boldsymbol{J}$ is electron magnetization flux density. To subsequent estimations, let us substitute into (7) the expression $\partial \boldsymbol{m}/\partial t = \omega\Delta\boldsymbol{m}$, where $\Delta\boldsymbol{m} = \boldsymbol{m} - \overline{\boldsymbol{m}}$, $\omega$ is an effective frequency of the vector $\boldsymbol{m}$ temporal oscillations.

According to Section 2, vector $\boldsymbol{m}$ perform a precession very quickly under $sd$ exchange field that is dominant in the range $0 \le x < \lambda_F$. The frequency of the precession is very high, namely, $\omega \sim \gamma\alpha M \equiv \omega_{sd}$. At typical parameter values: $\alpha \sim 2\times10^4$, $M \sim 10^3$ G, $\tau \sim 3\times10^{-13}$ s, the estimation $\omega_{sd}\tau \sim 10^2 \gg 1$ is valid. Therefore the last (relaxation) term in (7) can be neglected. Then we can calculate the products $(\Delta\boldsymbol{m}\boldsymbol{M})$ and $[\Delta\boldsymbol{m}, \boldsymbol{M}]$ from Eq. (7) and after this substitute the results back into Eq. (7). Finally we solve the equation with respect to $\Delta\boldsymbol{m}$ and get:

$$\Delta\boldsymbol{m} = -\omega_{sd}^{-1}\frac{\partial\boldsymbol{J}/\partial x + \zeta\big[\hat{\boldsymbol{M}}, \partial\boldsymbol{J}/\partial x\big] + \zeta^2\hat{\boldsymbol{M}}\big(\hat{\boldsymbol{M}} \cdot \partial\boldsymbol{J}/\partial x\big)}{1 + \zeta^2}, \tag{8}$$

where $\zeta = \omega_{sd}/\omega \sim 1$. It follows from (8), that the longitudinal and transversal (with respect to $M$) components of $\Delta m$ can be comparable in magnitude, that agrees with the picture of the electron spin precession in the range of the layer **2** near the interface (see Section 2).

However, a detailed analysis of the spin motion in this range is rather difficult, because flux $J$ is to be calculated and expressed via vector $m$. In the range $0 \leq x < \lambda_F$, it requires solving equation for density matrix, because the electrons are in quantum non-steady and inhomogeneous states. We try to avoid these difficulties further and show (see Section 4) that solving such a quantum problem is not necessary for our aims and can be replaced by introducing certain boundary condition.

Detailed analysis is to be carried out for the $x > \lambda_F$ range. In this range, vectors $m$ and $M$ are to be almost collinear. Correspondingly, the effective frequency $\omega$ is determined by the vector $M$ precession in rather low fields $H$, $\beta M$, $H_d << \alpha M$. We assume the condition

$$\omega \tau << 1 \tag{9}$$

is valid, which allows to neglect time derivative in (7) in comparison with the relaxation term. The expression (8) remains valid, but we should replace $\omega_{sd}^{-1} \to \tau$ and $\zeta \to \omega_{sd}\tau >> 1$. These replacements change all the estimations in (8). The last term in the numerator of (8) becomes dominant now and we get

$$\Delta m = -\tau \hat{M}\left(\hat{M} \cdot \partial J/\partial x\right). \tag{10}$$

By that, we confirmed the assumption made before (section 2) of an approximate collinearity of $\Delta m$ (and, consequently, the whole vector $m$) and $M$ vectors. The transversal components of $m$ still exist, however, but they are small as a parameter $(\omega_{sd}\tau)^{-1} \sim 10^{-2} << 1$ and will not be taken into account below.

The mobile electrons in the range $x > \lambda_F$, occupy two spin subbands with spin $\uparrow$ (parallel to $M$) and $\downarrow$ (antiparallel to $M$), respectively. Magnetization $m$ contains contributions from the both subbands and can be presented as

$$m = \mu_B\left(n_\uparrow - n_\downarrow\right)\hat{M} \equiv m\hat{M}, \tag{11}$$

where $\mu_B$ is Bohr magneton, $n_{\uparrow,\downarrow}$ are partial electron densities in the spin subbands. The total electron density $n = n_\uparrow + n_\downarrow$ does not depend on $x$ and $t$ because of local neutrality conditions in metal. The magnetization flux density $J$ in (10) also can be related with partial electric current densities in each subbands, $j_\uparrow$, and $j_\uparrow$. The relation has the following form

$$\boldsymbol{J} = \frac{\mu_B}{e}\left(j_\uparrow - j_\downarrow\right)\hat{\boldsymbol{M}}\,. \tag{12}$$

The total current density $j = j_\uparrow + j_\downarrow$ does not depend on $x$ and $t$ too, because of one-dimensional geometry of our model.

Let us consider a simple situation, when *sd* exchange gap in the layer **2** does not depend on $x$ and $t$ and, consequently, on the vector $\hat{\boldsymbol{M}}$ direction. Besides, the transport relaxation times in metals are rather small at room temperature, so that drift – diffusion approximation is valid. In scope of these assumptions, the spin current (12) has been calculated in [9]. By substituting the well known formulae for partial currents in (12) and following the way of calculations indicated in Ref. 9, we obtain

$$\boldsymbol{J} = \left(\frac{\mu_B}{e}Qj - \widetilde{D}\frac{\partial m}{\partial x}\right)\hat{\boldsymbol{M}}\,, \tag{13}$$

where $Q = \left(\sigma_\uparrow - \sigma_\downarrow\right)/\left(\sigma_\uparrow + \sigma_\downarrow\right)$ may be understood as the conductivity spin polarization parameter, and $\widetilde{D} = \left(\sigma_\uparrow D_\downarrow + \sigma_\downarrow D_\uparrow\right)/\left(\sigma_\uparrow + \sigma_\downarrow\right)$ is the spin diffusion constant, $\sigma_{\uparrow,\downarrow}$ and $D_{\uparrow,\downarrow}$ being partial conductivities and partial diffusion constants for spin up and spin down electrons, respectively. To obtain Eq. (13), an additional assumption should be made, namely,

$$\frac{j}{j_D} << 1\,, \tag{14}$$

where $j_D \equiv enl/\tau$ is a characteristic current density in the layer **2**. The condition (14) means that the current disturbs the subband populations rather slightly, so that a low injection level is realized. With typical parameter values, $n \sim 10^{22}$ cm$^{-3}$, $l \sim 3\times10^{-6}$ cm, $\tau \sim 3\times10^{-13}$ s, we get $j_D \sim 1.6\times10^{10}$ A/cm$^2$. Further in the paper, only much lower currents will be of interest, namely, $j \le 10^7 - 10^8$ A/cm$^2$, because the instability thresholds in study have just such an order of magnitude. Therefore, condition (14) is valid well in our calculations.

Now, let us substitute (13) into Eq. (10), which is a form of the equation of motion (7). Then we obtain

$$\frac{\partial^2 m}{\partial x^2} - \frac{m - \overline{m}}{l^2} = 0\,, \tag{15}$$

the $m(x,t)$ function being defined in (11), while its equilibrium value $\overline{m}$ was defined when discussion after the Eq. (7). Since the *sd* exchange gap is fixed, $\overline{m}$ value does not depend on $x$ and $t$. The spin diffusion length is $l = \sqrt{\widetilde{D}\tau}\,$.

## 4. Boundary conditions

To determine the necessary solutions of the dynamical equations (3) and (15), we derive boundary conditions for those equations. The derivation is based on the continuity condition for summary (mobile electrons and lattice) magnetization flux at the interfaces.

Since the lattice in the layer **1** is pinned, the magnetization flux $\boldsymbol{J}_1$ in the layer appears due to the electric current only and corresponds to spin transport of the mobile electrons. Therefore, the magnetization may be described by a formula similar to (13), but without the second term containing derivative with respect to $x$, because the mobile electron magnetization in the layer cannot be changed in space. Than the flux $\boldsymbol{J}_1$ takes the form

$$\boldsymbol{J}_1 = \frac{\mu_B}{e} Q_1 j \hat{\boldsymbol{M}}_1 ,\qquad(16)$$

where $\hat{\boldsymbol{M}}_1 = \boldsymbol{M}_1 / |\boldsymbol{M}_1|$, parameter $Q_1$ is defined analogously to $Q$ in (7) with partial conductivities and diffusion constants taken for the layer **1**. The flux (16) has both longitudinal and transversal components with respect to vector $\hat{\boldsymbol{M}}$, namely,

$$\boldsymbol{J}_{1\parallel} = \frac{\mu_B}{e} Q_1 j \left( \hat{\boldsymbol{M}}_1 \hat{\boldsymbol{M}} \right) \hat{\boldsymbol{M}} ,\qquad(17)$$

$$\boldsymbol{J}_{1\perp} = \frac{\mu_B}{e} Q_1 j \left[ \hat{\boldsymbol{M}}, \left[ \hat{\boldsymbol{M}}_1 , \hat{\boldsymbol{M}} \right] \right].\qquad(18)$$

Because of absence of longitudinal spin relaxation within the range $0 \le x < \lambda_F$, the longitudinal component of the flux (16) should be identical with the flux (13) near $x = 0$ interface. If we put $\hat{\boldsymbol{M}}_1 = -\hat{\boldsymbol{y}} \sin\chi + \hat{\boldsymbol{z}} \cos\chi$, where the unit vectors along the coordinate axes are $\hat{\boldsymbol{x}}, \hat{\boldsymbol{y}}, \hat{\boldsymbol{z}}$, then the following relation is to be fulfilled at $x = 0$:

$$\frac{\mu_B}{e} Q_1 j \left( -\hat{M}_y \sin\chi + \hat{M}_z \cos\chi \right) = \frac{\mu_B}{e} Q j - \widetilde{D} \frac{\partial m}{\partial x} .\qquad(19)$$

This is the first of desired boundary conditions.

The second boundary condition can be obtained by taking into account that the transversal component of the mobile electron magnetization flux disappears in the range $0 \le x < \lambda_F$ (see Eq. (2) and subsequent discussion). It follows from the total flux continuity that the flux (18) is to continue in the layer **2** as a lattice magnetization flux. The explicit expression for the latter one is derived in Appendix **I** (see formula (**I**.4)). Putting it equal to (18) at $x = 0$, we obtain

$$\frac{\mu_B}{e} Q_1 j \left[ \hat{\mathbf{M}}, \left[ \hat{\mathbf{M}}_1, \hat{\mathbf{M}} \right] \right] = aM \left[ \hat{\mathbf{M}}, \frac{\partial \hat{\mathbf{M}}}{\partial x} \right], \tag{20}$$

where $a = \gamma M A$, in accordance with Appendix **I**. The condition (19), in a slightly different form, was discussed and applied earlier (see, e.g., [9]). As to condition (20), it is, apparently, novel one. In the chosen coordinate system (see Fig. 1), the condition (20) gives two relations valid at $x = 0$:

$$\frac{\partial \hat{M}_x}{\partial x} + k\hat{M}_y \cos\chi + k\hat{M}_z \sin\chi = 0,$$
$$\frac{\partial \hat{M}_y}{\partial x} - k\hat{M}_x \cos\chi = 0, \tag{21}$$

where $k = \mu_B Q_1 j / eaM$. Component $\hat{M}_z$ can be found from $\hat{\mathbf{M}} = 1$ condition.

Now, let us consider the interface between the layers **2** and **3** at $x = L$. Layer **3** is nonmagnetic one, so that $Q_3 = 0$. Then the continuity condition for the longitudinal flux takes the form

$$\frac{\mu_B}{e} Q j - \widetilde{D} \frac{\partial m}{\partial x} = -\widetilde{D}_3 \frac{\partial m_3}{\partial x} \tag{22}$$

at $x = L$. Since the lattice magnetization is absent in the layer **3**, the lattice magnetization flux is to vanish at the interface, i.e., $aM \left[ \hat{\mathbf{M}}, \frac{\partial \hat{\mathbf{M}}}{\partial x} \right] = 0$, or, in components,

$$\frac{\partial \hat{M}_x}{\partial x} = 0, \quad \frac{\partial \hat{M}_y}{\partial x} = 0 \tag{23}$$

at $x = L$.

The next important conditions appear due to a penetration of mobile electrons through the interfaces. Such a penetration leads to the spin subband chemical potential differences are to be continuous at the interfaces. We neglect thermal spread of the Fermi steps in both subbands. Direct calculation leads in this case to the following expression for the chemical potential difference:

$$\zeta_\uparrow - \zeta_\downarrow = \left( 2\mu_B \right)^{-1} \left( \frac{1}{g_\uparrow(\overline{\zeta})} + \frac{1}{g_\downarrow(\overline{\zeta})} \right) \Delta m \equiv N\Delta m, \tag{24}$$

where $g_{\uparrow,\downarrow}(\zeta)$ are partial densities of states in the spin subbands, $\overline{\zeta}$ is an equilibrium chemical potential. The continuity condition for the difference (24) at the interface between the layers **2** and **3** may be written in the form

$$N\Delta m = N_3 \Delta m_3. \tag{25}$$

On the other hand, the chemical potential difference undergoes a break at the interface between the layers **1** and **2**. This is due to the idealization of the layer **1**, in which all the spins are assumed to be pinned completely.

We solve Eq. (15) with the boundary conditions (19), (22) and (25). Then we obtain the following distribution of the injected magnetization $\Delta m(x)$ in the layer **2**:

$$\Delta m(x) = \frac{j}{j_D} \frac{\mu_B n}{\sinh \lambda + \nu \cosh \lambda} \{Q \cosh \xi$$
$$+ \left[ Q_1 \left( -\hat{M}_y \sin \chi + \hat{M}_z \cos \chi \right) - Q \right] \cdot \left[ \cosh(\lambda - \xi) + \nu \sinh(\lambda - \xi) \right] \}, \tag{26}$$

where $\lambda = L/l$ and $\xi = x/l$. Parameter

$$\nu = \frac{nN}{n_3 N_3} \cdot \frac{j_{D3}}{j_D} \tag{27}$$

characterizes the influence of the layer **3**.

## 5. Static state and fluctuations

Let an external field **H** is applied along the positive direction of *z* axis, and the anisotropy axis is parallel to this axis, too. Then in absence of current, when there is no coupling between the layers in our model, the layer **2** magnetization is aligned along *z* axis.

With current turning on, spins flow through the interfaces $x = 0, L$. This current creates connection between **1**, **2** and **3** layers and can change the magnetization direction in the layer **2**. To describe this process, let us calculate $U_{sd}$ energy by substituting (26) into (6). We obtain

$$U_{sd} = -\alpha \overline{m} M L - \alpha (\hat{M}_1 \cdot M(+0)) l \cdot \mu_B n Q_1 \left( \frac{j}{j_D} \right) \cdot \left( 1 - \frac{\nu}{\sinh \lambda + \nu \cosh \lambda} \right). \tag{28}$$

To calculate variation derivative (5), $U_{sd}$ energy is to be presented as a functional of vector **M**(x). Then we obtain, in particular,

$$\frac{\delta}{\delta M(x)} (ML) = \frac{\delta}{\delta M(x)} \int_0^L M dx' = \int_0^L \hat{M}(x') \cdot \delta(x' - x) dx' = \hat{M}(x),$$

$$\frac{\delta}{\delta M(x)} \left( (\hat{M}_1 \cdot M(+0)) \right) = \hat{M}_1 \cdot \delta(x - \varepsilon), \tag{29}$$

with $\varepsilon \to +0$. By using (28) and (29), we calculate *sd* exchange field:

$$H_{sd}(x) = \alpha \overline{m} \hat{M}(x) + \alpha \hat{M}_1 \cdot \mu_B n Q_1 \left( \frac{j}{j_D} \right) \cdot \left( 1 - \frac{\nu}{\sinh \lambda + \nu \cosh \lambda} \right) \cdot l \delta(x - \varepsilon). \tag{30}$$

The first term in (30) is directed along **M** and does not contribute to the equation of motion (3). The second term is directed along **M₁**. Under an arbitrary χ angle, this term, as well as $H_{sd}$ field and effective field $H_{eff}$ as a whole (see (4)), may have direction that is different from the direction of *z* axis. It is clear, that the magnetization $\overline{M}$ in the static state, which is determined by $\left[\overline{M}, H_{eff}\right] = 0$ equation, may deviate from *z* axis, too. Besides, owing to the boundary conditions (21), $\overline{M}$ magnetization comes out to be inhomogeneous in the layer **2**. Once more great difficulty is that LLG equation determining $\overline{M}$ is nonlinear one.

Fortunately, two certain angles exist, namely χ = 0 and χ = π, for which vector $\overline{M}$ can be found easily. Indeed, at the angles indicated, (21) and (23) conditions fulfil for the vector $\overline{M}$ that is independent on coordinate *x* and has components $\overline{M}_x = 0$ and $\overline{M}_y = 0$. Since vector $\overline{M}$ is parallel to *z* axis, the equation of motion (3) also fulfils in the static state. Therefore, we obtain for χ = 0, π

$$\overline{M}_x = \overline{M}_y = 0, \qquad \overline{M}_z = M > 0. \tag{31}$$

We consider further only these last indicated values of the angle χ and therefore investigate the stability only for the static state (31). In particular, we have $\boldsymbol{M_1} = M_1 \cdot \hat{\boldsymbol{z}}$ with $M_1 = \cos \pi \cdot |\boldsymbol{M_1}| = -|\boldsymbol{M_1}| < 0$ at χ = π.

It is convenient for subsequent calculations to replace δ-function in the formula (30) with the following finite expression:

$$\delta(x - \varepsilon) \rightarrow \frac{1}{r} \cdot \exp\left(-\frac{x}{r}\right) \tag{32}$$

with subsequent passing to the limit $r \rightarrow +0$. Under such a passage, the right-hand side of (32) is nonzero only at *x* = 0 point, while integral of the function over *x* from 0 to ∞ is equal to 1.

Let us introduce fluctuations Δ**M** by the following way:

$$\boldsymbol{M} = M \cdot \hat{\boldsymbol{z}} + \Delta \boldsymbol{M}. \tag{33}$$

We linearize LLG equation (3) with respect to Δ**M** and than substitute into that equation the exchange field in the form (30) and the demagnetization field in the form $\boldsymbol{H}_d = -4\pi \Delta M_x \cdot \hat{\boldsymbol{x}}$. After the substitution we obtain

$$\frac{\partial \Delta M_x}{\partial t} + \kappa \frac{\partial \Delta M_y}{\partial t} = a \frac{\partial^2 \Delta M_y}{\partial x^2} - \left(\Omega_y - \gamma B \frac{l}{r} \exp\left(-\frac{x}{r}\right)\right) \Delta M_y,$$

$$\frac{\partial \Delta M_y}{\partial t} - \kappa \frac{\partial \Delta M_x}{\partial t} = -a \frac{\partial^2 \Delta M_x}{\partial x^2} + \left(\Omega_x - \gamma B \frac{l}{r} \exp\left(-\frac{x}{r}\right)\right) \Delta M_x,$$

$$\tag{34}$$

where the following notations are used:

$$\Omega_x = \gamma(H + H_a + 4\pi M), \qquad \Omega_y = \gamma(H + H_a), \tag{35}$$

$$B = \alpha \frac{\mu_B jl}{e\widetilde{D}} \cdot Q_1 \left(1 - \frac{\nu}{\sinh\lambda + \nu\cosh\lambda}\right), \tag{36}$$

$H_a = \beta M$ is anisotropy field. The coefficients in (34) contain $x$ variable. It is convenient to change the variable:

$$y = b \cdot \exp\left(-\frac{x}{2r}\right), \qquad b = 2\sqrt{\gamma Blr/a} \ . \tag{37}$$

Then the temporal Fourier components $\Delta M_x$, $\Delta M_y \sim \exp(-i\omega t)$ obey the following equations:

$$y^2 \frac{\partial^2 \Delta M_x}{\partial y^2} + y\frac{\partial \Delta M_x}{\partial y} + \left(y^2 - \frac{4r^2\Omega_x}{a} + \frac{4r^2}{a} i\kappa\omega\right) \cdot \Delta M_x = \frac{4r^2}{a} i\omega \cdot \Delta M_y,$$

$$y^2 \frac{\partial^2 \Delta M_y}{\partial y^2} + y\frac{\partial \Delta M_y}{\partial y} + \left(y^2 - \frac{4r^2\Omega_y}{a} + \frac{4r^2}{a} i\kappa\omega\right) \cdot \Delta M_y = -\frac{4r^2}{a} i\omega \cdot \Delta M_x. \tag{38}$$

Equations (38) are familiar in structure to the Bessel equation, however, the components $\Delta M_x$ and $\Delta M_y$ in the equations are coupled with each other. To uncouple the components, consider a set of equations

$$\left[s^2 - \frac{4r^2}{a}(\Omega_x - i\kappa\omega)\right] \cdot \Delta M_x - \frac{4r^2}{a} i\omega \cdot \Delta M_y = 0,$$

$$\left[s^2 - \frac{4r^2}{a}(\Omega_y - i\kappa\omega)\right] \cdot \Delta M_y + \frac{4r^2}{a} i\omega \cdot \Delta M_x = 0. \tag{39}$$

The solvability condition for the set gives the following expression for $s^2$:

$$s^2 = \frac{2r^2}{a} \cdot \left[\Omega_x + \Omega_y - 2i\kappa\omega \pm \sqrt{(\Omega_x - \Omega_y)^2 + 4\omega^2}\right]. \tag{40}$$

As seen, there are two different values of $s^2$. Each of them corresponds to a certain fluctuation type and is to be taken into account in stability analysis. Let us take any of the solutions (40), substitute it to Eq. (39) and use the equations to exclude $\Delta M_y$ component from the first of Eqs. (38) and $\Delta M_x$ component from the second one. As a result, each of the Eq. (38) transforms into a standard equation for the Bessel functions with $s$ index. Therefore, general solution of the Eq. (38) can be written as

$$\Delta M_{x,y}(y) = A_{x,y} \cdot J_s(y) + B_{x,y} \cdot Y_s(y), \tag{41}$$

where $J_s$ and $Y_s$ are Bessel functions of the first and the second kind, respectively [22], $A_{x,y}$ and $B_{x,y}$ are the constants to be found from the boundary conditions (21) and (23) taken at $\chi = \pi$. The solubility condition from which four constants $A_{x,y}$ and $B_{x,y}$ are to be found is written down in Appendix **II**. There the limiting passage prescribed by substitution (32) is carried out. As a result, the following dispersion relation is obtained for fluctuations:

$$qL \cdot \tan qL = \Phi , \qquad (42)$$

where

$$\Phi = -\frac{\gamma B L^2}{a\lambda} \pm ikL , \qquad (43)$$

and parameter $q \equiv is/2r$ is determined by (40) and further plays a role of a wavenumber.

The form of expression (42) coincides with the well-known dispersion relation for standing spin waves in a layer of ferromagnet [20, 23]. The right-hand side $\Phi$ has a meaning of an effective pinning parameter for the lattice spins at the layer faces (at $x = 0$ plane, in particular). Note that in our model there is no induced magnetic anisotropy pinning the surface lattice spins. It may be seen immediately from the boundary conditions (21) and (23). Nevertheless, an effective pinning takes place due to the current $j$. According to (43), the real part of $\Phi$ describes the pinning due to the exchange effective field, while the imaginary one describes effective pinning due to the torque. It is interesting, in this connection, that an experimental studying of spin-wave resonance (SWR) absorption in magnetic junctions could give an important information about the surface spin pinning character.

The dispersion relation (42) is used further to calculate complex eigenfrequency $\omega$ of the fluctuations and find the instability conditions, that are the conditions when $\operatorname{Im}\omega \geq 0$ .

## 6. Instability of fluctuations

Let us analyze the solutions of Eq. (42). The simplest situation corresponds to the absence of any current. By definition of parameters $B$ and $k$ (see (36) and (21)), it follows from (43) that $\Phi = 0$ at $j = 0$. The case corresponds to the SWR under free (unpinned) spins at the layer sides. In this case, Eq. (42) has the following solutions:

$$qL = n\pi \text{ and } \omega = \sqrt{\left(\Omega_x + a\frac{n^2\pi^2}{L^2}\right) \cdot \left(\Omega_y + a\frac{n^2\pi^2}{L^2}\right)} - \frac{i\kappa}{2}\left(\Omega_x + \Omega_y + 2a\frac{n^2\pi^2}{L^2}\right), \quad (44)$$

$n = 0, \pm1, \pm2,...$ Formulae (44) are obtained with using definition $q = is/2r$ and Eq. (40). Since $\operatorname{Im}\omega < 0$ in (44), the fluctuations are stable due to dissipation in absence of current.

Let us turn on now the current with $j/e > 0$ . According to (43), we have

$$\operatorname{Im}\Phi/\operatorname{Re}\Phi = \kappa\eta << 1, \qquad (45)$$

where

$$\eta \equiv \left[ \alpha \gamma M \tau \kappa \left( 1 - \frac{\nu}{\sinh \lambda + \nu \cosh \lambda} \right) \right]^{-1}. \qquad (46)$$

Parameter (46) may be less or more than 1, or it may be of the order of 1. As it will be seen somewhat further, the parameter describes relation between the torque and spin injection contributions to the current induced instability threshold and increment. But the inequality (45) typically remains valid due to low value of the dissipation parameter $\kappa \ll 1$.

Let us estimate the right-hand part of (42) taking (45) into account. We have

$$|\Phi| \approx \text{Re} \, \Phi = \lambda \left( 1 - \frac{\nu}{\sinh \lambda + \nu \cosh \lambda} \right) \cdot \theta^{-1}(j), \qquad (47)$$

where $\theta(j) \equiv eAM/\alpha \mu_B j \tau Q_1 \, l$ is current dependent. Take, for example, typical parameter values $\alpha \sim 2 \times 10^4$, $Q_1 \sim 0.3$, $\tau \sim 3 \times 10^{-13}$ s, $l \sim 17$ nm, $A \sim 10^{-12}$ cm$^2$, $M \sim 10^3$ G. At $j \sim 3.3 \times 10^7$ A/cm$^2$, that corresponds to the instability threshold (see, e.g., [1-3]), we have $\theta(j) \approx 0.17$. Figure 2 allows find the values $\lambda = \lambda'$ at which $|\Phi| = 1$ under the different $\nu$ indicated. As seen, $0.4 < \lambda' < 0.7$ that corresponds to $7 \leq L \leq 12$ nm. As far as we know, only thicknesses $L$ within the range from 2 to 10 nm was studied so far [1]. In accordance with the estimates, inequalities $|\Phi| \ll 1$ or $|\Phi| \leq 1$ fulfil at such thicknesses. However, at somewhat larger thicknesses $L \approx 20 - 80$ nm (or $\lambda \approx 1 - 5$) $|\Phi| \gg 1$ condition can fulfil. Therefore, we consider both large and small values of $|\Phi|$ (really, the limiting cases $|\Phi| \ll 1$ and $|\Phi| \gg 1$).

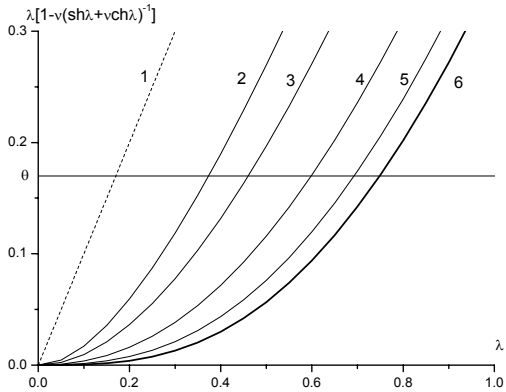

Fig. 2. A diagram to determine junction parameters, which correspond to various values of the effective lattice spins pinning parameter $|\Phi|$: $|\Phi| = 1$ at $\theta(j)$ level, $|\Phi| < 1$ below this level, and $|\Phi| > 1$ above it. The $\nu$ parameter takes the following values: 1 - $\nu = 0$; 2 - $\nu = 0.5$; 3 - $\nu = 1$; 4 - $\nu = 3$; 5 - $\nu = 10$; 6 - $\nu = \infty$.

6.1. Instability at $|\Phi| \ll 1$. The solutions of Eq. (42) are close to SWR modes (44). First, consider a solution close to homogeneous resonance with $n = 0$. We retain only the main term $q^2 L^2 \ll 1$ in the left-hand side of Eq. (42) and use the definition $q = is/2r$, as well as formulae (40), (43). Then we obtain a quadratic equation with respect to frequency $\omega$:

$$\omega^2 = \left[ \Omega_x - \frac{\gamma B}{\lambda} - i\left( \kappa\omega \pm \frac{ak}{L} \right) \right] \cdot \left[ \Omega_y - \frac{\gamma B}{\lambda} - i\left( \kappa\omega \pm \frac{ak}{L} \right) \right]. \tag{48}$$

The roots of the equation are complex ones. The direct calculations of the real and imaginary parts of the roots give

$$\left(1 + \kappa^2\right) \cdot \operatorname{Re}\omega = \pm \frac{\kappa ak}{L} \pm \frac{1}{\sqrt{2}} \cdot \sqrt{W + \sqrt{W^2 + V^2}} \ , \tag{49}$$

$$\left(1 + \kappa^2\right) \cdot \operatorname{Im}\omega = -\frac{\kappa}{2}\left( \Omega_x + \Omega_y - \frac{2\gamma B}{\lambda} \right) \pm \frac{1}{\sqrt{2}} \cdot \sqrt{-W + \sqrt{W^2 + V^2}} \ , \tag{50}$$

where any combinations of the signs may be taken, and

$$W = \left( \Omega_x - \frac{\gamma B}{\lambda} \right) \cdot \left( \Omega_y - \frac{\gamma B}{\lambda} \right) - \frac{(ak)^2}{L^2} - \frac{\kappa^2}{4} \cdot \left( \Omega_x - \Omega_y \right)^2,$$

$$V = \frac{ak}{L} \cdot \left( \Omega_x + \Omega_y - \frac{2\gamma B}{\lambda} \right). \tag{51}$$

We retain only the upper sign in (50), which allows instability. Then the instability condition takes the form

$$\kappa^2 \cdot \left( \Omega_x - \frac{\gamma B}{\lambda} \right) \cdot \left( \Omega_y - \frac{\gamma B}{\lambda} \right) - \left( \frac{ak}{L} \right)^2 \le 0 \ . \tag{52}$$

Two sources of instability occurrence are seen clearly from the condition (52). One of them is the injected spin contribution to the effective field (30). This contribution is described with the terms proportional to parameter $B$ (see (36)). The other instability source is the torque contribution acting at the boundary $x = 0$. The last contribution arises due to the boundary conditions (21) and is described with the term $\sim k^2$.

According to (52), the dissipation loss influences two contributions indicated by principally different ways. Dissipation is not substantial for injection type instability, because $\gamma B/\lambda > \Omega_y$ condition is sufficient. On the other hand, if only the torque is retained ($B \to 0$), then some threshold due to dissipation is to be overcome. At the same time, joint action of both mechanisms always decreases the left-hand part of (52) and, consequently, facilitates occurrence of instability.

Since $B$ and $k$ are proportional to current, we can obtain a quadratic equation determining the threshold current density $j_{th}$ by equating the right-hand part of (52) to zero. Solving the equation gives

$$\frac{j_{th}}{j_\perp} = \eta \cdot \frac{\left(f + f^{-1}\right) - \sqrt{\left(f - f^{-1}\right)^2 + 4\eta^2}}{2\left(1 - \eta^2\right)}, \tag{53}$$

where

$$j_\perp = \frac{\kappa e M \omega_0 L}{\mu_B Q_1}, \tag{54}$$

$\omega_0 = \sqrt{\Omega_x \Omega_y}$ means SWR frequency at $n = 0$ (cf. (44)), $f = \sqrt{\Omega_y / \Omega_x}$. The parameter $j_\perp$ (54) is nothing but the threshold current density under only the torque acting, i. e., at $B = 0$. It follows from the general formula (53) for the threshold, that $j_{th} \to j_\perp$ at $\eta \gg 1$. On the other hand, if $\eta \ll 1$, then the spin injection instability mechanism dominates. In the latter case, the threshold is

$$j_{th} = j_\perp \cdot f\eta = j_D \cdot \frac{(H + H_a)\lambda(\sinh\lambda + \nu\cosh\lambda)}{\alpha\mu_B n Q_1 [\sinh\lambda + \nu(\cosh\lambda - 1)]}, \tag{55}$$

that coincides with one obtained earlier [9] by other method, namely, by using minimal magnetic energy condition (remember the current density $j_D$ is defined in (14)). In complete correspondence with the paper [9], formula (55) refers to homogeneous fluctuations, because the SWR mode with $qL \ll 1$ is taken.

Note an important relationship between threshold currents and SWR frequencies. It follows from (49) and (50)

$$\frac{(\mathrm{Re}\,\omega)_{th}}{\omega_0} = \frac{j_{th}}{j_\perp}. \tag{56}$$

Ratio (56) tends to zero at $\eta \to 0$ and to 1 at $\eta \to \infty$. This is because the torque can influence only the SWR damping and can change its sign at the instability threshold. Torque do not change the resonance frequency. On the other hand, injection influences also the real part of the frequency by making it vanish at the threshold. The eigenfrequency softening occurs, which is analogous with a reorientation transition in an external magnetic field. This allows an interesting possibility to identify experimentally the injection mechanism by measuring the resonance frequency at the current density near the threshold value.

Ratio (56) as a function of $\eta$ parameter at various $f$ values is shown in Fig. 3. Note that the instability threshold is always lower than the torque induced threshold (54). Let us estimate the threshold (54) using the set of typical parameters indicated after formula (47). Additionally, let us take: damping constant $\kappa = 3 \times 10^{-2}$, resonant frequency $\omega_0 = 2.3 \times 10^{10}$ c$^{-1}$, and $\lambda = 0.4$. Then we obtain $j_\perp \approx 2.7 \times 10^7$ A/cm$^2$. A similar estimate is given by formula (55) within its validity range. Therefore, estimates of the threshold current density agree well with experimental data.

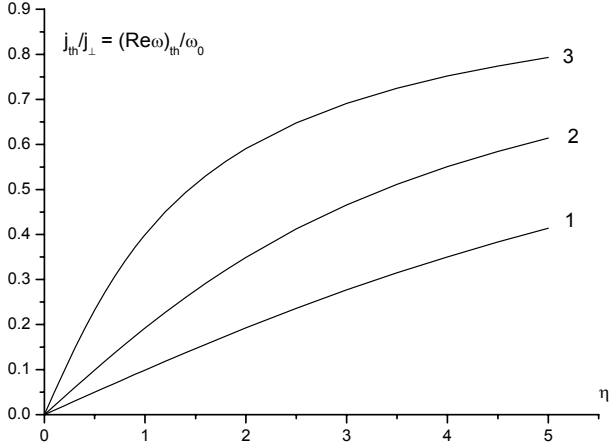

Fig. 3. The threshold current density and spin wave eigenfrequency at the threshold as functions of parameter $\eta$ that characterizes relative contributions of the torque and spin injection. The curves differ each other in the values of parameter $f = \sqrt{\Omega_y / \Omega_x}$ : $1 - f = 0.1$; $2 - f = 0.2$; $3 - f = 0.5$.

6.2. Instability of the inhomogeneous SWR modes having $n \neq 0$ numbers at $|\Phi| << 1$. Corresponding solutions of Eq. (42) are close to the zeros of $\tan qL$, i.e.,

$$q \approx \frac{n\pi}{L} - \frac{\Phi}{n\pi L}. \tag{57}$$

Using formula (40) once more and definitions $q = is/2r$ and (43), we have a quadratic equation to find frequency $\omega$. We separate real and imaginary parts of the equation roots, as it has been done above for Eq. (48). Then the instability condition $\mathrm{Im}\,\omega > 0$ takes the form

$$\kappa^2 \cdot \left[ \Omega_x + a\left(\frac{n\pi}{L}\right)^2 - \frac{2\gamma B}{\lambda} \right] \cdot \left[ \Omega_y + a\left(\frac{n\pi}{L}\right)^2 - \frac{2\gamma B}{\lambda} \right] - \left(\frac{2ak}{L}\right)^2 < 0 \tag{59}$$

which is similar to that of Eq. (52). Therefore, the conclusions about the effective field and torque contributions retain validity.

The threshold current density $j_{th}^n$ with $n \neq 0$ is calculated just the same way as in the preceding section. We obtain then

$$j_{th}^n = \frac{1}{2} j_\perp \cdot \frac{\eta}{2(1-\eta^2)} \cdot \left[ 2n^2 g + f + f^{-1} - \sqrt{(f^{-1} - f)^2 + 4(1 + n^2 g f)(1 + n^2 g f^{-1})\eta^2} \right], \tag{59}$$

where $g = \pi^2 a / \omega_0 L^2$ parameter is introduced. It follows from (59) that the instability threshold rises abruptly with number $n$.

6.3. Now consider an instability at $|\Phi| \gg 1$. In that case, the dispersion relation (42) is greatly modified by the current. Because of (45) inequality, there is a solution with dominant imaginary part of $q$, as can be confirmed by a direct substitution. For the solution,

$$q^2 \approx -(\Phi/L)^2 . \qquad (60)$$

By using $q = is/2r$ and (43) definitions, we transform Eq. (42) in a quadratic equation for frequency $\omega$:

$$\omega^2 = \left[ \Omega_x - \frac{(\gamma Bl)^2}{a} + ak^2 - i(\kappa\omega \pm 2\gamma Blk) \right] \cdot \left[ \Omega_y - \frac{(\gamma Bl)^2}{a} + ak^2 - i(\kappa\omega \pm 2\gamma Blk) \right] . \qquad (61)$$

Equation (61) differs significantly from (48) in two aspects. First, the current induced corrections to frequencies $\Omega_{x,y}$ are quadratic in the current, rather than linear. Secondly, both mechanisms, injection field ($\sim B$) and torque ($\sim k^2$), contribute simultaneously to the real and imaginary parts of the multipliers in the square brackets in Eq. 61.

The instability condition $\operatorname{Im}\omega > 0$ takes the form

$$\kappa^2 \cdot \left[ \Omega_x - \frac{(\gamma Bl)^2}{a} + ak^2 \right] \cdot \left[ \Omega_y - \frac{(\gamma Bl)^2}{a} + ak^2 \right] - (2\gamma Blk)^2 < 0 . \qquad (62)$$

As distinct from the condition (52), the instability becomes impossible if the injection effective field is absent, i.e., at $B = 0$. On the hand, in absence of the torque ($k = 0$), the instability may be possible under $(\gamma Bl)^2 / a > \Omega_y$ condition.

Since $B \sim j$ and $k \sim j$, we obtain a biquadratic equation for determining the threshold current density. By solving the equation, we obtain

$$j_{th} = \frac{e}{\mu_B \alpha \gamma \tau Q_1} \cdot \sqrt{\frac{a}{2(1 - 4\eta^2)}} \cdot \sqrt{\Omega_x + \Omega_y - \sqrt{(\Omega_x - \Omega_y)^2 + 16\eta^2 \Omega_x \Omega_y}} . \qquad (63)$$

This formula is valid under the condition $|\Phi| \gg 1$. The condition corresponds to the range that lies well above the θ level in Fig. 2. In this range, all the curves rise linearly with λ. Then parameter η does not depend, practically, on λ (i.e., on the layer thickness $L$), in accordance with (46). Therefore the threshold (63) does not depend on $L$, too. Such a feature contrasts with the linear rising of the threshold at $|\Phi| \ll 1$ (see (54) and (55)). The difference is due to the fact, that the inhomogeneous fluctuations with $|q| \gg L^{-1}$ are most unstable at $|\Phi| \gg 1$, in accordance with the solution (60). Using a standard set of parameters, we may find easy the thickness $L \geq 20$ nm should be taken to realize the condition $|\Phi| \gg 1$. Then the threshold current density becomes $j_{th} \approx 3 \times 10^6$ A/cm$^2$ at $\Omega_x \approx 2.3 \times 10^{11}$ s$^{-1}$, $\Omega_y \approx 1.8 \times 10^9$ s$^{-1}$, so that the threshold may be lower by an order of magnitude in comparison with the estimates made using formulae (54) and (55), which are valid for a very small $L$.

## 7. Electric current effect on spin wave spectra

The current through the junction influences not only the spin wave decrement, $\mathrm{Im}\,\omega$, but also the spin wave spectrum, $\mathrm{Re}\,\omega$. We discuss such an influence in more detail by considering two effects, as examples: 1) softening of the spin modes near the instability threshold, and 2) appearance of new spin modes due to the current.

7.1. We consider the softening effect for $n = 0$ mode with neglecting dissipation. Let us assume $|\Phi| \ll 1$, for definiteness. We obtain then immediately from Eq. (48)

$$\frac{\mathrm{Re}\,\omega}{\omega_0} = \sqrt{\left(1 - \frac{f}{\eta} \cdot \frac{j}{j_\perp}\right) \cdot \left(1 - \frac{1}{f\eta} \cdot \frac{j}{j_\perp}\right)}. \tag{64}$$

By using (53), we have finally

$$\frac{\mathrm{Re}\,\omega}{\omega_0} = \left\{ \left[ 1 - \frac{f\left[ f^{-1} + f - \sqrt{\left(f^{-1} - f\right)^2 + 4\eta^2} \right]}{2\left(1 - \eta^2\right)} \frac{j}{j_{th}} \right] \left[ 1 - \frac{f^{-1} + f - \sqrt{\left(f^{-1} - f\right)^2 + 4\eta^2}}{2f\left(1 - \eta^2\right)} \frac{j}{j_{th}} \right] \right\}^{1/2}. \tag{65}$$

A number of curves are plotted in Fig. 4 using formula (65). The threshold $j_{th}$ in the formula should be taken at the corresponding value of $\eta$ for each curve. One can see that $\mathrm{Re}\,\omega$ decreases with approaching the instability threshold. The decrease becomes steeper with decreasing $\eta$, i.e., under conditions, when the injection effective field becomes more important.

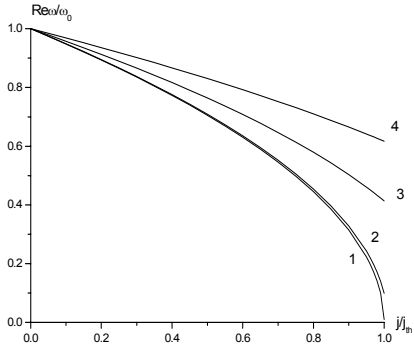

Fig. 4. The spin wave eigenfrequency as a function of the current density. The frequency softening occurs with approaching the threshold. The results correspond to $n = 0$ mode at $|\Phi| \ll 1$ and $f = 0.1$. The curves differ in the values of parameter $\eta$: 1 – $\eta = 0.1$; 2 – $\eta = 1$; 3 – $\eta = 5$; $\eta = 10$.

In the opposite case, $|\Phi| \gg 1$, similar curves was obtained also ,and the frequency decrease near the threshold appears more abrupt.

7.2. The mentioned effect of new mode appearance takes place at $|\Phi| \gg 1$. A number of solutions of Eq. (42) having the real part of $q$ much more than imaginary one appears in that limiting case. These solutions exist additionally to the only solution considered in section 6.3. Since the condition (45) always fulfils, Eq. (42) takes the form

$$qL \cdot \mathrm{tg}\, qL \approx \mathrm{Re}\, \Phi \gg 1. \qquad (66)$$

The solutions mentioned are close to

$$q \approx \frac{\pi}{L} \cdot \left( n + \frac{1}{2} \right). \qquad (67)$$

According to SWR theory [20, 23], such solutions with half-integer numbers in brackets correspond to natural oscillation of a layer with tightly pinned surface magnetization. In our case, the pinning arises due to large value of $\mathrm{Re}\, \Phi$, i.e., to high injection effective field. The situation looks like as if the injected spins create unidirectional magnetic anisotropy near the surface $x = 0$ due to $sd$ exchange; it is the anisotropy that pins the magnetization.

## 8. Conclusions

The joint action of two electric current induced effects is considered for the first time in the present work, namely, 1) nonequilibrium spin injection, and 2) surface torque. These effects act together on a state of a ferromagnetic free layer of spin-valve type metallic junction.

At the current densities exceeding certain threshold (typically, $10^6$ - $3 \times 10^7$ A/cm$^2$) spin-wave fluctuations in the junction become unstable. The joint action of the effects mentioned lowers the instability threshold.

The surface torque determines the instability threshold in junctions with thin enough free layer ($L \sim 2-7$ nm) and low dissipation ($\kappa < 10^{-2}$). In this case, the instability threshold rises linearly with $L$ and corresponds, practically, to homogeneous excitation of the ferromagnetic layer.

The nonequilibrium longitudinal spin injection determines the instability threshold in junctions with thick enough free layer ($L > 20$ nm) and high enough dissipation ($\kappa \sim 3 \times 10^{-2}$). Under such conditions, the threshold corresponds to inhomogeneous excitation of the layer. The threshold current density appears to be low enough, $\sim 10^6$ A/cm$^2$.

Current induced injection of nonequilibrium longitudinal spins causes instability because of reorientation phase transition in the exchange effective field created by the spins. Therefore, the injection influences not only the fluctuation decrement, but also the fluctuation spectrum.

The eigenfrequencies of the spin fluctuations tend to zero (soften) with the current increasing and approaching to the threshold value.

In contrast, the surface torque influences only the decrement, not the spin fluctuation spectrum. This fact can be used to identify the instability mechanism in experiment.

Under high spin injection in ferromagnetic metallic layer, effective pinning occurs of the layer magnetization near the injecting contact. As calculations show, such a pinning leads to appearance of new SWR modes under current increasing.

The authors would like to thank A.I. Krikunov, A.V. Medved, and V.A. Sablikov for interesting discussions. This work was supported by the Russian Foundation for Basic Research (Grants No. 03-02-17540 and 04-02-08248).

## Appendix **I**
## The lattice magnetization flux

The LLG equation takes the form

$$\frac{\partial \boldsymbol{M}}{\partial t} = -\gamma A \cdot \left[ \boldsymbol{M}, \frac{\partial^2 \boldsymbol{M}}{\partial x^2} \right] - \gamma [\boldsymbol{M}, \boldsymbol{H}_{s-d}] - \gamma [\boldsymbol{M}, \boldsymbol{H}'] + \kappa \left[ \hat{\boldsymbol{M}}, \frac{\partial \boldsymbol{M}}{\partial t} \right], \tag{I.1}$$

where $\boldsymbol{H}' \equiv \boldsymbol{H} + \beta(\boldsymbol{Mn})\boldsymbol{n} + \boldsymbol{H}_d$. Let us assume the estimates $\kappa << 1$ and $\left| [\boldsymbol{M}, \boldsymbol{H}'] \right| << \left| [\boldsymbol{M}, \boldsymbol{H}_{s-d}] \right|$ to be valid within $0 \le x < \lambda_F$ range. Then two last terms in the right-hand side of (**I**.1) can be neglected. Transform the vector product with the second derivative in (**I**.1) by the following way:

$$\left[ \boldsymbol{M}, \frac{\partial^2 \boldsymbol{M}}{\partial x^2} \right] = \frac{\partial}{\partial x} \left[ \boldsymbol{M}, \frac{\partial \boldsymbol{M}}{\partial x} \right] - \left[ \frac{\partial \boldsymbol{M}}{\partial x}, \frac{\partial \boldsymbol{M}}{\partial x} \right] = \frac{\partial}{\partial x} \left[ \boldsymbol{M}, \frac{\partial \boldsymbol{M}}{\partial x} \right]. \tag{I.2}$$

With simplifications made and taking into account (**I**.2), the Eq. (**I**.1) takes the form

$$\frac{\partial \boldsymbol{M}}{\partial t} + \gamma [\boldsymbol{M}, \boldsymbol{H}_{s-d}] + \frac{\partial}{\partial x} \left\{ a \left[ \hat{\boldsymbol{M}}, \frac{\partial \boldsymbol{M}}{\partial x} \right] \right\} = 0, \tag{I.3}$$

where $a = \gamma A M$.

The first term in (**I**.3) is the magnetization change in time, while the second one is proportional to the torque, which the mobile electrons affect the lattice. Hence, the last term is a divergence of the lattice magnetization flux. Therefore,

$$\boldsymbol{J}_\perp = a \left[ \hat{\boldsymbol{M}}, \frac{\partial \boldsymbol{M}}{\partial x} \right] \tag{I.4}$$

means the lattice magnetization flux, while $a$ is magnetization diffusion constant.

## Appendix **II**

## Towards the derivation of the dispersion relation for fluctuations

Substitution of the general solution (41) into the boundary conditions (21) and (23) leads to the following solvability condition:

$$I_1^2 + 4k^2 I_2^2 = 0 , \tag{II.1}$$

where

$$
\begin{aligned}
I_1 &= J_s{}'(b) \cdot Y_s{}'\!\left(be^{-\frac{L}{2r}}\right) - Y_s{}'(b) \cdot J_s{}'\!\left(be^{-\frac{L}{2r}}\right), \\
I_2 &= \frac{r}{b}\left[ J_s(b) \cdot Y_s{}'\!\left(be^{-\frac{L}{2r}}\right) - Y_s(b) \cdot J_s{}'\!\left(be^{-\frac{L}{2r}}\right)\right],
\end{aligned}
\tag{II.2}
$$

the primes mean differentiating with respect to $y$ variable. According to (32), we should pass to the limit $r \to +0$. It means, in particular, a limiting case should be considered of small indices and arguments of the Bessel functions, because of $b \sim \sqrt{r}$ and $s \sim r$, in accordance with (37) and (40). We use the following approximate formulae:

$$J_s(y) \approx \frac{y^s}{2^s\,\Gamma(s+1)} \cdot \left[1 - \frac{y^2}{4(s+1)}\right], \tag{II.3}$$

$$Y_s(y) \approx \frac{1}{\pi s} \cdot \left[\frac{y^s}{2^s\,\Gamma(s+1)} - \frac{y^{-s}}{2^{-s}\,\Gamma(-s+1)}\right] \cdot \left(1 - \frac{y^2}{4}\right). \tag{II.4}$$

We take these functions and their derivatives at $y = b$ and $y = be^{-\frac{L}{2r}}$. Substituting the results into (**II**.2) and retaining the main terms at $r \to 0$ gives

$$
\begin{aligned}
I_1 &\approx \frac{2}{\pi} e^{\frac{L}{2r}} \frac{s}{b^2} \cdot \left(\sinh\frac{Ls}{2r} - \frac{b^2}{2r}\cosh\frac{Ls}{2r}\right), \\
I_2 &\approx \frac{2}{\pi} e^{\frac{L}{2r}} \frac{r}{b^2} \cdot \cosh\frac{Ls}{2r}.
\end{aligned}
\tag{II.5}
$$

We use formulae $\Gamma(s+1) = s\Gamma(s)$ and $\Gamma(s)\Gamma(-s) = -\dfrac{\pi}{s \cdot \sin \pi s} \approx -\dfrac{1}{s^2}$ here.

By substituting (**II**.5) in (**II**.1) we obtain

$$\left(\tan\frac{Ls}{2r} - \frac{b^2}{2s}\right)^2 + \left(\frac{2kr}{s}\right)^2 = 0 . \tag{II.6}$$

The ratios $s/r$ and $b^2/s$ tend to finite limits at $r \to 0$. With $q = is/2r$ notation and the parameter $b$ definition (37) taking into account, the dispersion relation (42) is obtained.